\documentclass[manuscript]{aastex}
\usepackage{graphicx}
\usepackage{epstopdf}

\shorttitle{HZ-Her: New Radius and Distance}
\shortauthors{Leahy and Abdallah}

\begin{document}

\title{HZ Her: Stellar Radius from X-ray Eclipse Observations, Evolutionary State and a New Distance}


\author{D. A. Leahy and M.H. Abdallah}
\affil{Dept. of Physics, University of Calgary, University of Calgary,
Calgary, Alberta, Canada T2N 1N4}



\begin{abstract}
Observations of HZ Her/Her X-1 by the Rossi X-ray Timing Explorer (RXTE)
covering high state eclipses of the neutron star are analyzed here. 
Models of the eclipse are used to measure the radius and atmospheric scale height
of HZ Her, the stellar companion to the neutron star. 
The radius is  2.58 to 3.01 $\times10^{11}$ cm, depending on system inclination and mass ratio(q), 
with accuracy of $\sim$1 part in 1000 for given inclination and $q$.
We fit Kurucz model stellar atmosphere models to archival optical observations. 
The resulting effective temperature ($T_{eff}$) of the unheated face of HZ Her is determined to be 
in the 2$\sigma$ range 7720K to 7865K, and 
 metallicity ($log(Z/Z_{\odot})$) in the range -0.27 to +.03. 
The model atmosphere surface flux and new radius yield a new
distance to HZ Her/ Her X-1, depending on system inclination and $q$:
a best-fit value of 6.1 kpc with upper and lower limits of 5.7 kpc and 7.0 kpc.
We calculate stellar evolution models for the range of allowed masses (from 
orbital parameters), and allowed metallicities (from  optical spectrum fits).
The stellar models agree with $T_{eff}$ and radius of HZ Her for two narrow
ranges of mass: 2.15 to 2.20 $M_{\odot}$ and  2.35 to 2.45 $M_{\odot}$
This lower mass range implies a low neutron star mass (1.3 $M_{\odot}$),
whereas the higher mass range implies a high neutron star mass (1.5-1.7 $M_{\odot}$).

\end{abstract}



 \section{Introduction}

Her X-1/HZ Her is a bright and well-studied persistant X-ray binary pulsar
(e.g. Klochkov et al. 2009, Ji et al. 2009, 
Leahy 2003, Leahy 2002, Igna 2010, Crosa \& Boynton 1980, Gerend \& Boynton 1976).
 The system is located at a distance of approximately 6 kpc from Earth. 
 It consists of an A7 type stellar companion (HZ Her) which varies between late A and early B with the
orbital phase, and a neutron star (Her X-1) with masses 
$\sim$2.2 $M_\odot$ and $\sim$1.5 $M_\odot$, respectively (Reynolds et al. 1997). 
An updated set of binary parameters is given by \citet{stau09}. 
It is one of the few neutron star binaries to have low interstellar absorption, 
which makes it feasible for observation and study at various wavelengths 
(e.g. Leahy 2003, Leahy 2002, Scott et al. 2000, Scott \& Leahy 1999). 
It is additionally of special interest because it is the neutron star with the best-known
mass-to-radius ratio \citep{lea04b}.
This value was made possible by a definitive determination of the geometry of the emitting
region on the neutron star from a study of the 35-day evolution of the pulse profile \citep{sco00}.
Thus it of great interest to constrain the mass of the neutron star better than the
current range of $\simeq$1.3-1.5$M_{\odot}$.
This would allow the determination of both mass and radius with tight constraints and 
form a strong constraint on equations of state for dense matter in neutron stars.

Emission from Her X-1/HZ Her covers the optical, ultraviolet, EUV and
X-ray regime and models for the, often coupled, emission processes
must ultimately be consistent.
X-ray spectra of Her X-1 are discussed by \citet{oos98} and
\citet{dal98} (from BeppoSAX), \citet{cho97} (from ASCA) and \citet{lea95b} (from GINGA). 
The hard X-rays ($>1$ keV) arise as a result of 
mass accretion onto the neutron star and are modulated by the neutron star 
rotation and obscuration by the accretion disk (e.g. Scott et al. 2000 and 
references therein), companion star and cold gas in the system that causes 
the absorption dips (Leahy 1997, Igna and Leahy 2011, Igna and Leahy 2012).
A small reflected/reprocessed X-ray component is present that is observable during the low state and eclipses (Choi et al. 1994, Leahy et al. 1995).
A major portion of the observed optical/ultraviolet emission is believed to 
arise from X-ray heating of HZ Her and the accretion disk. 
The X-ray heating causes the surface temperature of HZ Her facing the neutron star to be 
approximately 10,000 degrees higher than the cooler shadowed side \citep{cheng95}. 
Analysis of ultraviolet spectra of Her X-1 are presented by \citet{bor97} and \citet{vrt96}. 
Optical signatures of reprocessing on the companion and accretion disk are discussed by \citet{sti97}.
Observations of the broad band optical emission of HZ Her/Her X-1
have been presented by \citet{dee76} and \citet{vol90}, among others. 
Between the hard X-ray and 
optical/ultraviolet band lies the soft X-ray/extreme ultraviolet band ($\sim0.016$ $-$ 1 keV). 
A blackbody spectral component has been detected by 
many previous observations with a temperature of about 0.1 keV 
(e.g. Shulman et al. 1975, Oosterbroek et al. 2000)
and generally has been attributed to 
reprocessing of hard X-rays in the inner region of the accretion disk 
(e.g. McCray et al. 1982, Oosterbroek et al. 2000). 

Her X-1 shows a variety of phenomena at different timescales, including 1.24 second pulsations, orbital eclipses 
with period of 1.7 days, and a 35-day cycle in the X-ray flux. 
The latter normally consists of a bright state, known as the Main High state, which lasts 10-12 days and an intermediate-brightness state, the Short High state, which lasts 5-7 days. 
These two states are separated by 8-10 day long low states.
The precessing accretion disk causes the alternating pattern of High and Low X-ray intensity 
states by periodically blocking the neutron star from view. 
The initial rapid flux rise over a few hours marks the start of a 
High state and is known as the ``turn-on''. 
This event is the emergence of the 
neutron star from behind the moving outer disk edge (see \citet{sco00} and references therein). 
Discussions of the properties of the 35-day cycle are given by
\citet{lea10}, \citet{sco99} and \citet{sha98}. 
The X-ray pulse profile evolution has been convincingly linked to the precessing accretion disk by \citet{sco00}. 
Over the course of the 35-day cycle, the broadband optical emission exhibits a complex, systematic 
variation, in addition to the orbital modultation due to X-ray heating of HZ Her. 
This pattern is a consequence of disk emission and of disk shadowing/occultation 
of the heated face of HZ Her by the precessing accretion disk \citep{ger76}. 

The regular variability in Her X-1 that is of interest here is X-ray eclipse of the
neutron star during Main High. 
During Main High the observer has a direct view of the neutron star, so that eclipse
probes the structure of the companion star HZ Her.
The previous study of this nature was carried out by \citet{lea95} using one eclipse
observed with GINGA.
Here we consider all Main High eclipses from the RXTE/PCA data archive, covering more than 
one decade of observations of Her X-1.  
Accurate measurements of the radius of HZ Her are made and combined with constraints from
optical spectra of HZ Her during eclipse, which yield effective temperature ($T_{eff}$) 
and metallicity (Z), to obtain a new distance to the system.
We compare the observationally determined radius and effective temperature to 
stellar evolution calculations to determine allowed masses for HZ Her, then use
those to explore the allowed parameters of the system, including the mass of the neutron star.

\section{RXTE Observations of Eclipses and Analysis}

\subsection{Eclipse Light Curves}

\begin{figure}
\epsscale{.80}
\plotone{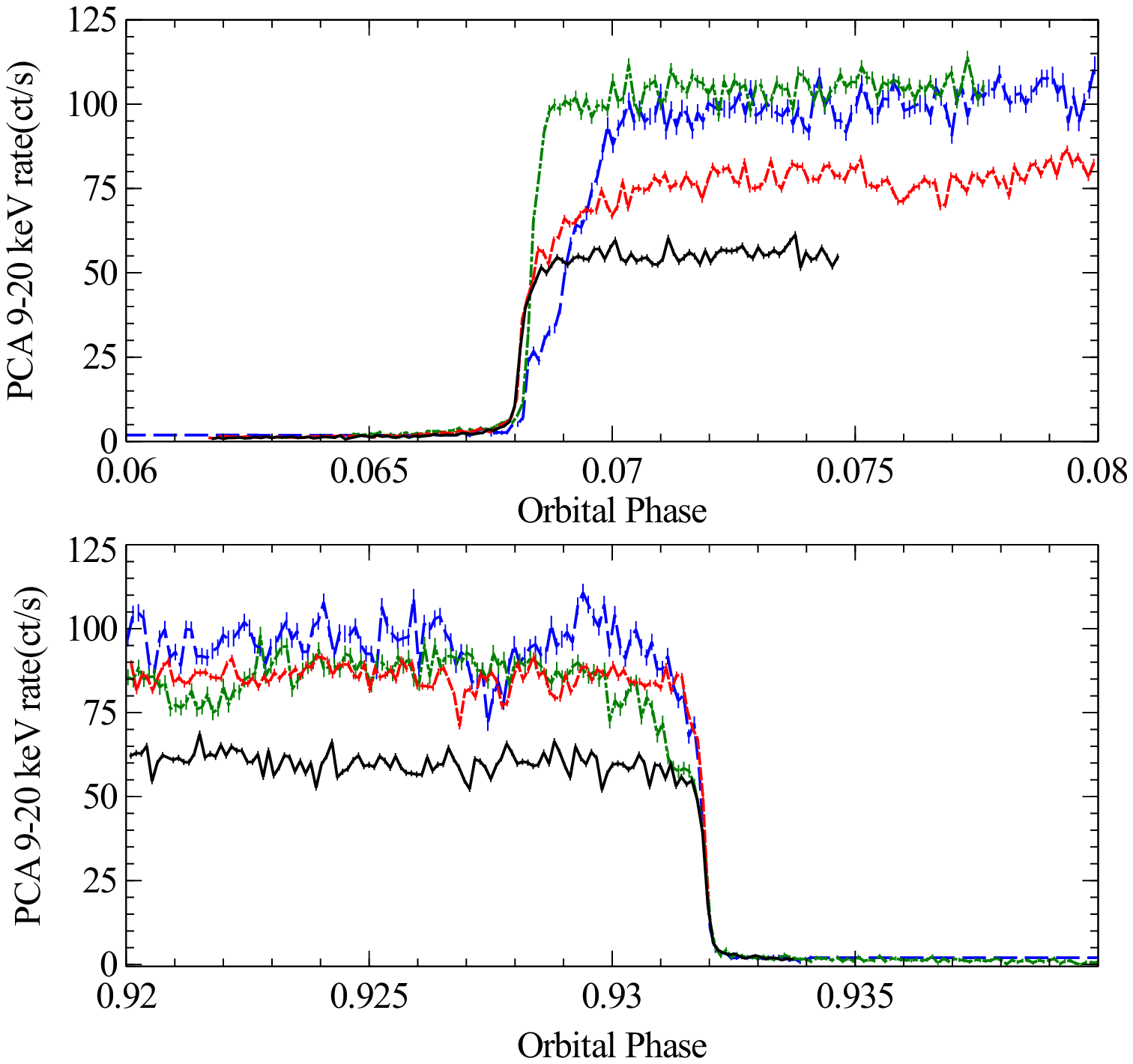}
\caption{High State eclipses without excess absorption observed with RXTE/PCA: the four egresses (top panel: MJD50357- black, solid line; MJD50361- red, dash line; MJD52597- green, dash-dot line; MJD52599- blue, long-dash line) and four ingresses(lower panel: MJD50361- black, solid line; MJD50710- red, dash line; MJD52599- green, dash-dot line; MJD52601- blue, long-dash line).\label{fig1}}
\end{figure}

A summary analysis of the entire RXTE/PCA set of Standard 2 data is given in \citet{lea11}. 
We are here interested in analyzing Main High ingresses or egresses to obtain values
of the radius of HZ Her.
These eclipses are known to occur rapidly (over a few minutes) at orbital phases $\simeq$0.93 
and $\simeq$0.07 (e.g. Leahy and Yoshida 1995)
and are caused by the limb of HZ Her covering
the observer's line-of-sight to the neutron star.
The part of the data which occurs during Main High state was determined here using the Main High
turn-on times of \citet{lea10}.
The Main High data was, in turn, searched for all data during orbital phases of eclipse
(including ingress and egress: phase 0.92 to 1.08). 
Nineteen different eclipses were observed with varying amounts of phase coverage. 
Counting the ingress part (phase 0.92 to 1.0) and egress part (phase 0.0 to 0.08) separately, 
there were 23 ingresses or egresses during Main High state. 
Some of these were observed with too little phase coverage to measure the ingress or egress light curve.
Several other ingresses or egresses include cold matter
absorption immediately around the time of ingress or egress.
This is seen from the low (2-4 keV)/(9-20 keV) softness ratio coinciding with below-normal count rates, which indicates absorption. 
Sporadic absorption events (dips) are regularly seen during main high state. This has been
recently quantified by \cite{lea11}: their Fig. 5 shows the data for main high state folded on the
orbital period, with absorption dips seen as drops in count rate simultaneous with a decrease in the
(2-4 keV)/(9-20 keV) softness ratio; their Fig. 7 shows the fraction of time spent in dips to the total
time observed as a function of orbital phase in main high state. Around time of eclipse, this fraction is
between 0.3 and 0.6.
There were four clear (without excess absorption) ingresses and four clear egresses suitable for analysis.
Figure~\ref{fig1} shows the RXTE (9-20keV) light curves for these eight ingresses/egresses and 
Table~\ref{tbl-data} gives the MJD, 35 day phase and orbital phase coverage of these observations.

\subsection{Orbital Parameters of HZ Her/Her X-1}

To model the system orbit we use the most recent published orbital parameters, given by 
\citet{stau09}. 
These parameters are summarized here in Table~\ref{tbl-orbit}.
To calculate the absolute size of the orbit, the mass ratio is needed, which can be found
from the amplitude of the velocity of the companion star HZ Her, $K_c$, 
together with the amplitude of the velocity of the neutron star, $K_x$ (from $q=M_x/M_c=K_c/K_x$).
Because $K_x$ is already very well known through pulse time delays, a good value for $K_c$ is needed. 
The best-fit value \citep{rey97} is $K_c$=109 km/s  with a quoted statistical
error of 3 km/s and systematic error of $\sim$10 km/s. 
Here, in addition to the best-fit case, we take a low $K_c$ case of  99 km/s and 
a high $K_c$ case of 119 km/s value.
A range of inclinations between $80^\circ$ and  $90^\circ$ was chosen, which is somewhat
larger than normally adopted inclinations ($\simeq 83^\circ$ to  $\simeq 87^\circ$) 
to explain the 35 day cycle (see e.g. Leahy 2002).
The inclination and $K_c$ values significantly affect the orbit size and the masses of the 
two components HZ Her and Her X-1, as shown in  Table~\ref{tbl-mass}; 
errors in the masses and size of the orbit have
a negligible contribution from the other measured orbital parameters. 
 
\subsection{Eclipse Light Curve Model}

A numerical model was constructed and fitted to the observed light-curves to 
determine the radius of HZ Her. 
The duration of eclipse in orbital phase is determined by the amount of angle, relative to 360$^\circ$ of
a full rotation of the binary system, during which HZ Her blocks the line-of-sight to the neutron star.
The conversion of this angle into physical size depends on the binary separation, a.
The conversion of this size into a stellar radius depends on the binary inclination, i, as the line-of-sight
makes a different cut across the back face of the star depending on inclination.

A few approximations are made in the model used here.
The eccentricity of the orbit is very small (see Table~\ref{tbl-orbit}), enough that use a circular orbit gives
accurate results (to 1 part in $10^4$).
The atmosphere of the star was taken to have an exponential density profile, and the radius of the
star was defined as the radius where the vertical optical depth was unity in the 9-20 keV X-ray
band. 
For the 9-20 keV band, the absorption plus scattering opacity was taken to be sum of
the photoelectric and Thompson cross-sections, using matter of solar composition.
The neutron star is taken as a point source of X-rays.
The accretion column that emits the bulk of the X-rays is small enough ($<$a few km, 
Leahy 2004, Leahy 2004b)
that a point source is a good assumption for the X-ray source. 
A few km is small compared to the next smallest scale involved, which is the scale height of 
the atmosphere of HZ Her ($\sim6\times10^8$cm, Leahy and Yoshida 1995). 
HZ Her is essentially Roche-lobe filling (e.g. Leahy and Scott 1998), 
so it is not spherical but has the shape of the Roche lobe. 

The model initially uses a spherical star, and yields the impact parameter from the center of the 
star to the line-of-sight through the stellar surface at vertical optical depth unity. 
We call this distance $R_{tan}(\tau=1)$ (or for simplicity, just $R_{tan}$).
Knowing the shape of the Roche lobe, which depends only on $q$, we use $R_{tan}$ to calculate
the distance to any other part of the surface on the Roche lobe.

We fit the model to each of the eight observed ingresses and egresses. 
The MJD52599 egress (Fig.1 top panel) shows a extended slow rise after orbital phase 0.068: 
the fit for this egress was restricted to the early sharp rising part of the egress. 
For the MJD52599 ingress (Fig.1 bottom panel) there is a shallow decrease prior to the
main sharp ingress- the fit was restricted to the sharp part of the ingress. 
The eight ingresses/egresses were fit separately is to demonstrate that
a single radius can explain all events, so that there is no evidence that the radius of HZ Her 
is time-variable. 
The resulting radii ($R_{tan}$) and scaleheights (H) are given in Table ~\ref{tbl-eclipse85} 
for the case of $85^{\circ}$ inclination and $K_c$=109 km/s.
The uncertainty of $R_{tan}$ is $\sim 10^{8}$ cm and of H is  $\sim  10^{8}$ cm (i.e. similar 
in absolute magnitude, but the relative uncertainty in $R_{tan}$ is three orders of magnitude smaller).
We find that for a given inclination and $K_c$, all eight eclipses give the same $R_{tan}$ within errors. 
This confirms we are measuring the radius of HZ Her and not some other variable phenomenon, 
e.g. related to the accretion disk or stream. 
As a result, we average $R_{tan}$ over the eight eclipses for each inclination and $K_c$ 
and present the results in Table \ref{tbl-eclipseav}. 

For all of the eclipses, inclinations and $K_c$ values, $H$ is consistent with a constant value of
$5\times 10^{8}$cm, consistent with the value found by \citet{lea95}.
The uncertainty in $R_{tan}$ here is estimated from the root-mean-square of the eight 
best-fit values of $R_{tan}$ for each inclination at $K_c$ value, which ranges from  
$.0023\times 10^{11}$ to $.0031\times 10^{11}$ cm for the fifteen different cases.

Next we calculate a correction for the non-spherical shape of the surface of the star, which
is filling its Roche lobe, using the standard Roche potential. 
The radius derived for a spherical star, $R_{tan}$ (listed in Table~\ref{tbl-eclipseav}), is 
from the center of HZ Her, to the point on the Roche lobe surface 
tangent to the line-of-sight from observer to neutron star.
We will need the cross-sectional area of the star viewed by the observer:
this will be used in Sect. 4.1 together with the surface flux to
derive the distance to the binary. 
For the fifteen different cases (five inclinations and three mass ratios- see Table~\ref{tbl-mass})
$R_{tan}$ ranged from 0.412 to 0.441 times the orbital separation. 
To determine the cross-sectional area of the star at mid-eclipse, because the system inclination is near
90$^\circ$, we can use the area of an ellipse with
major-axis radius $R_{tan}$, and minor axis the polar radius of the Roche Lobe of HZ Her.
We find the polar radius ranges from 0.943 to 0.948 times $R_{tan}$. 
Thus to accuracy of about one part in 400, we use a fixed ratio of 0.9455 for all cases,
which gives a cross-sectional area during eclipse of $0.9455\pi R_{tan}^2$.

For comparison with stellar models, one needs the radius of a sphere, $R_L$, with the same volume 
as the distorted star. 
In the case of HZ Her the star just fills its Roche lobe, so we can
use the formula of \citet{egg83} which relates $R_L/a$ to mass ratio, $q$. 
For the three different cases of mass ratio here (see Table~\ref{tbl-mass}) one obtains 
$R_L/a=$ 0.426, 0.417 and 0.410, respectively.
To determine the radius of the spherical star of the same volume, the $R_{tan}$
radii in Table~\ref{tbl-eclipseav} need to be multiplied by the ratio of  $R_L/R_{tan}$,
which depends on mass ratio, $q$, and inclination.
For $q$=0.586, this ratio ranges from to 1.0047 ($80^\circ$ inclination) to 1.0025 ($90^\circ$),
while for $q$=0.645, the range is 1.0060 to 1.0041, 
and for $q$=0.704, it is 1.0071 to 1.0052. 
Thus to one part in 400, this ratio can be take as 1.005 for all cases here.
However, we used the more accurate numerically calculated value of the ratio to obtain $R_L$ 
for each case of $q$ and inclination separately. 

\section{Constraints on Effective Temperature and Metallicity of HZ Her}

During eclipse of the neutron star, only the unheated surface of HZ Her is viewed from Earth. 
The optical spectrum during eclipse has been used to estimate the luminosity
and thus distance of HZ Her/Her X-1 previously (e.g. Oke 1976, Cheng et al. 1995).
Here we compare moderately high resolution optical observations to 
the detailed set of stellar atmosphere spectra computed by \citet{Kur93} over a comprehensive
grid in $T_{eff}$ and log($Z/Z_{\odot}$) and with log(g) set to the value appropriate for HZ Her. 

High resolution optical spectra of HZ Her during orbital eclipse were published by \citet{and94}, 
and were taken with the {\it Hubble Space Telescope} Faint Object Spectrograph (FOS) (for the wavelength ranges $\sim 1200-1600 \mathring{A}$ and $\sim 2230-3230 \mathring{A}$) 
and with the {\it Kitt Peak National Observatory} GoldCam instrument 
(for the wavelength ranges $\sim 3200-6000 \mathring{A}$). 
However, as noted by \citet{cheng95}, the GoldCam spectrum is not flux calibrated, so
is not suitable for use in quantitative modeling.
Earlier spectrophotometry for the wavelength range $\sim 3000-10000 \mathring{A}$
was published by \citet{oke76} for orbital phase 0.011 (hereafter referred to as Oke data):  
those data have absolute flux calibration and are suitable for use here.
The GoldCam data was directly compared to the Oke data. It is found that the GoldCam data is 
slightly different in shape but noticibly higher (15\%) in flux- much larger than the data errors. 
This confirms the lack of flux calibration of the GoldCam data.
The HST/FOS, GoldCam and Oke data sets are not simultaneous, but the back side of HZ Her 
should not be variable.

The HST/FOS data covers two wavelength ranges (2220  to 3200$\mathring{A}$  and 1285 to 1604$\mathring{A}$)  
and were taken with resolution $(\lambda / \Delta \lambda)$ $\simeq$1300. 
The GoldCam data covers the range 3500$\mathring{A}$ to 6000$\mathring{A}$, with $3.9\mathring{A}$ resolution. 
We digitized the HST/FOS and GoldCam data at orbital phase 0 from Figures 2c and 2d of \citet{and94} 
at resolutions of $\sim$2$\mathring{A}$ and $4\mathring{A}$, respectively.  
The data errors were estimated from the Figures.  
The short wavelength HST/FOS data (1285$\mathring{A}$ to 1604$\mathring{A}$)  from Figure 2b of that paper
were not used for final analysis because they are 
dominated by the line emission from hot gas (e.g. Anderson 1994, Cheng 1995) and not HZ Her. 
The Oke data were digitized and the data errors were estimated from the Table 2 of \citet{oke76}.
Those data (covering 3280 to 9880$\mathring{A}$) have a bandpass of 40$\mathring{A}$ for wavelengths 
less than 5800$\mathring{A}$ and 80$\mathring{A}$ for wavelengths greater than 5800$\mathring{A}$,
with a few wavelength bins missing.
The combined data set used here for analysis consisted of the Oke data plus the HST/FOS 2220 to 3200$\mathring{A}$
data.

The Kurucz stellar atmosphere models are available as fits files 
from the {\it Space Telescope Science Institute}  ({\it STScI})website.
They have spectral resolution (bin size) of $10\mathring{A}$ over the range of  
$1200-3000 \mathring{A}$ and  $20\mathring{A}$ above  $3000 \mathring{A}$.
Prior to carrying out any spectral fits,
the HST/FOS data was rebinned the data to match the bins for the lower-resolution 
Kurucz model atmospheres in the range 2220 to 3200$\mathring{A}$. 
The Kurucz model atmosphere data was rebinned to match the bins for lower-resolution Oke data
in the range 3280 to 9880$\mathring{A}$.
This resulted in 183 wavelength bins for the observations and the matching Kurucz models.
Fitting was based on least-squares ($\chi^2$) minimization. To be conservative and to allow for
some systematic errors, the estimated errors of the HST/FOS and Oke data sets were multiplied by
a factor of 1.8. This does not affect the best fits but makes the  2 and 3 $\sigma$ limits on
$T_{eff}$ and log($Z/Z_{\odot}$) obtained from the spectral fits somewhat larger (about a factor of 2)
than if no allowance is made for an additional source of error. 

The resulting allowed mass range of HZ Her is $\simeq$2.2 to 2.5 $M_{\odot}$ 
(see Table~\ref{tbl-mass}) with higher values for lower inclinations.
The stellar radius from the eclipse fits gives higher values for lower inclinations, resulting
in a calculated surface gravity of log(g) nearly equal 3.50 for all inclinations.
Thus we fixed log(g)=3.50 for all stellar atmosphere models, leaving $T_{eff}$ 
and log($Z/Z_{\odot}$) as free parameters. 

We carried out fits of stellar atmosphere spectra for a wide range of $T_{eff}$ 
and log($Z/Z_{\odot}$). The results are presented in  Table~\ref{tbl-spectra}.
The minimum $\chi^2$ per degree-of-freedom (182, since these fits have one free parameter, normalization)
is $\simeq$1, indicating that our error estimates are reasonable.
These fit showed that $T_{eff}$ was near the center of the range 7250-8250K,  
and indicated that a much better constraint on $T_{eff}$ was achievable.
We obtain a new approximate constraint on metallicity, which is refined below:  
log($Z/Z_{\odot}$) is in the range -1.0
to +0.2. Outside of the range, the observed spectrum strongly disagrees with the short-wavelength
observations and the strength of the Balmer jump at $\sim 3800 \mathring{A}$.

Next we carried out fits with a finer grid of stellar models with $T_{eff}$ in the range 
7500-8000K and log($Z/Z_{\odot}$) in the range -0.5 to +0.2. To obtain a finer resolution in $T_{eff}$ 
than the archived spectra from {\it STScI} (which was 250K in $T_{eff}$), we interpolated in 
the Kurucz models in $T_{eff}$. 
The result of these spectral fits are summarized in Table~\ref{tbl-spectra}.
The 2 and 3 $\sigma$ limits for 2 parameters of interest ($T_{eff}$ and log($Z/Z_{\odot}$)) are 
$\delta\chi^2$=6.8 and 11.8, respectively. 
We fit with a finer grid of models around the best-fit model to find
the allowed region in the $T_{eff}$ vs. log($Z/Z_{\odot}$) plane.
The limits of this region are given in  Table~\ref{tbl-TZlimits}.
The extreme edges of the 2-$\sigma$ allowed region extend to 7720 K$<T_{eff}<$7865 K and 
$-0.27<log(Z/Z_{\odot} <0.03$.
The best fit model and the observations are shown in Figure~\ref{fig3}.

\section{Discussion}

\subsection{A New Distance to HZ Her}

We can directly calculate a new distance to HZ Her. 
The radius of HZ Her is determined from the X-ray eclipses, and the 
surface flux is determined from the stellar atmosphere model fits to HZ Her.
The observed continuum surface flux during eclipse was found by \citet{sti97}
and is shown in Fig. 2 (top panel) of that paper, obtained from the line-free wavelength
 ranges 4745-4755$\mathring{A}$, 4775-4796$\mathring{A}$ and 4898-4912$\mathring{A}$.
 During mid-eclipse the flux is 6.5 mJy, with an error of $\sim$0.2-0.3mJy.
The Kurucz model atmosphere spectra have 20 $\mathring{A}$ bins in this wavelength region, so we chose 
only line-free bin 4780-4800 $\mathring{A}$ in the model spectra to obtain the model surface continuum flux.
The error in distance caused by the uncertainty in continuum flux is $\sim$2\%, small compared to the
uncertainties in distance from other factors (inclination and $T_{eff}$).

We include in the distance calculation the correction for the shape of the Roche lobe, as 
 described above.
The resulting distances are given in Table~\ref{tbl-distance}.
The distances are the best-fit values and the larger (smaller) of the two
 upper (lower) limits to distance from the 2$\sigma$ $T_{eff}$ and log($Z/Z_{\odot}$) 
 limits determined above. 
 Because the measured stellar radius, $R_{tan}$ from the X-ray eclipses depends on inclination,
the best-fit distance and upper and lower limits depend on inclination. 

Our results are about 5\% smaller than the distance derived by \citet{rey97}, which is
mainly attributed to our $T_{eff}$ of $\simeq7800$K, which is lower than the value
of $\simeq8100$K they used from \citet{cheng95}. 

\subsection{Stellar Models, Mass and Evolutionary State of HZ Her}

Stellar evolution models were obtained using the EZ-Web interface (which can be found at http://www.astro.wisc.edu/~townsend/) 
to obtain a basic understanding of the evolutionary state of HZ Her. 
The code itself is an updated version of the 
of the \citet{egg71} code, which includes several improvements 
including the ability to evolve past the Helium flash in low mass stars.
In this simplified treatment we do not take into account effects of the
binary on the stellar evolution, even though there is evidence of mass
transfer from the progenitor of Her X-1 onto HZ Her from the enhanced N abundance
of HZ Her \citep{jim005}.
A detailed study of stellar evolution is beyond the scope of this paper.
Even so, the available standard stellar evolutionary models in the mass range
of interest (2-2.5 $M_{\odot}$) are sufficiently reliable to infer the properties of HZ Her.

The mass range of interest is 2.139 to 2.586 $M_{\odot}$ (see Table~\ref{tbl-mass})
for an inclination range of $80^\circ$ to $90^\circ$.
The metallicity range of interest are the best fit and upper and lower 2$\sigma$ limits,
given in Table~\ref{tbl-TZlimits}.  
Using $Z_{\odot}$=0.02, the best fit and upper and lower limits on $Z$ are 0.015, 0.011 and 0.021, respectively.
Models with $Z=0.01$ and 0.02 were computed to beyond the point where they begin 
ascent up the giant branch, which happens from $\sim$600 Myr to 1 Gyr after zero-age 
main sequence, depending on initial mass.
The examples plotted here are for $Z=Z_{\odot}=0.02$.
The evolutionary tracks for stars of 2.2, 2.3, 2.4 and 2.5 $M_{\odot}$ are shown in
Figure~\ref{fig4} for the region of interest here:
from zero-age main sequence until they become too cool 
to possibly represent the current stage of HZ Her, which has $T_{eff} \simeq$ 7800K. 

We next compare the evolutionary tracks with the limits on effective temperature
$T_{eff}$ and stellar radius $R$.
The $T_{eff}$ limits come from the spectral fits to the optical spectrum of HZ Her (discussed
 above) and are independent of mass of the star. 
To show allowed the region for HZ Her in the HR diagram, the 2-$\sigma$ upper and lower limits
 to $T_{eff}$ are shown in Figure~\ref{fig4} by the vertical dotted lines.
The  values of $R$ derived from the X-ray eclipses   
 depend both on inclination and $K_c$ (see Table~\ref{tbl-eclipseav}).
However one can take the global upper and lower limits of $R$ from Table~\ref{tbl-eclipseav} 
to calculate global upper and lower limits to the luminosity of HZ Her using
$L=4\pi R^2 \sigma_{SB} T_{eff}^4$ with $\sigma_{SB}$ the Stefan-Boltzmann constant.
These temperature-dependent upper and lower limits are plotted on Figure~\ref{fig4} as
the sloping dotted lines.

The measured stellar radius $R$ has negligible uncertainty other than that from the uncertainty in inclination and $K_c$.
We note that each different evolutionary track is essentially a constant mass track because
 the mass loss rates are very small prior to ascending the giant branch. 
The result is that, with fixed mass for each evolutionary track, $K_c$ and inclination are
no longer independent variables. 
We can take $K_c$ to be a function of inclination. 

Thus in order to obtain $R$ limits for each mass, we allow inclination to vary over its full
range ($80^\circ$ to $90^\circ$) but reject cases which have $K_c$ outside its allowed
range of 99 to 119 km/s. 
A separate calculation is done for each input value of companion mass $M_c$. 
Table~\ref{tbl-massnew} gives the allowed range of inclinations and associated $K_c$ for a set of
fixed masses for HZ Her. 
Note that the range of allowed inclination (or $K_c$) is very small 
near the mass limits (e.g. for 2.15 and 2.55 $M_{\odot}$).
Also given are the semi-major axis, $a$, the neutron star mass, $M_x$, and
the eclipse derived radius. The radius at the
minimum inclination is larger than that for the maximum inclination, which happens because the
line-of-sight from observer to neutron star makes a cut across the stellar surface at higher latitude
at lower inclination.
From Table~\ref{tbl-massnew} it is seen that the upper and lower limits to stellar radius, $R$, are
significantly different for different masses of HZ Her. 

The net result is that each evolutionary track has it own $R$ limits and (common) $T_{eff}$ limits.
Figure~\ref{fig5} shows example evolutionary tracks in the $R$ vs. $T_{eff}$ plane for M=2.22 
and 2.25 $M_{\odot}$ and the $R$ limits and  $T_{eff}$ limits for these two cases. 
As time increase the tracks move up and to the left in the diagram, then turn to move down and
to the right (a brief time of increasing temperature and decreasing radius), and then move upward
and left as the star becomes a red giant.
For both 2.22 and 2.25 $M_{\odot}$ cases, the evolutionary tracks cross their respective 
allowed regions only during
the short phase of increasing temperature and decreasing radius just prior to central hydrogen exhaustion.

We explore a range of masses. 
Below 2.15  $M_{\odot}$, the evolutionary track does not pass into the allowed $R$-$T_{eff}$ region
but below and left of it.
Between 2.15 and 2.20 $M_{\odot}$, the point of central hydrogen exhaustion is inside the allowed
region, and the duration that the evolutionary track is in the allowed region is short 
($\sim$100kyr).
Between 2.20 and 2.35 $M_{\odot}$, the situation is like that for 2.22 and 2.25 $M_{\odot}$ cases,
where the evolutionary track only crosses the allowed region during the phase of increasing temperature and decreasing radius.
Between 2.35 and 2.45 $M_{\odot}$, the evolutionary track crosses the allowed region before the rightward
excursion in the $R$ vs. $T_{eff}$ diagram, with decreasing temperature and increasing radius 
(before the leftward turn in the H-R diagram, see Figure~\ref{fig4}).
Masses above 2.45 $M_{\odot}$ do not cross the allowed region in the $R$ vs. $T_{eff}$ diagram:
all three parts of the evolutionary track pass above (too large $R$) and to the right (too
large $T_{eff}$) of the allowed region.

The requirement that the evolutionary track passes through the allowed $R$-$T_{eff}$ region, gives
a maximum mass of 2.45 $M_{\odot}$. 
If we make the sensible requirement that HZ Her is in a phase of expanding radius, to drive the
mass accretion onto the neutron star, then the mass range 2.20 to 2.35 $M_{\odot}$ is not
allowed. 
This leaves two allowed mass ranges.
For 2.15 to 2.20 $M_{\odot}$, the track is inside the allowed region and has increasing radius
just after the point of central hydrogen exhaustion. 
The rate of radius expansion from the stellar models is rapid, $\simeq1\times10^{-3}$cm/s, 
and the duration that the track is inside the allowed region is very short.
For 2.35 to 2.45 $M_{\odot}$, the track is inside the allowed region with increasing radius. 
The rate of radius expansion is slower, $\simeq 2.5-3.5\times10^{-5}$cm/s, 
but the duration that the track is inside the allowed region is longer 
($\sim$1-2Myr for 2.35 and 2.45
$M_{\odot}$ and $\sim$10Myr for 2.40 $M_{\odot}$, where the track crosses the center of the allowed
region).

We take a mass (2.40 $M_{\odot}$) near the centre of the higher mass allowed region 
as an example case to illustrate the evolutionary state of HZ Her.
The star enters the allowed region for HZ Her in the $R$-$T_{eff}$ plane at an age of 616 Myr. 
It leaves the allowed region at an age of 628 Myr. 
Shortly thereafter the radius continues to expand, then it decreases for a short time just 
before the age of 666 Myr. 
At this time the radius starts to rapidly increase, while
at the same time, the surface temperature changes from increasing to decreasing again. 
This point is marked by central H exhaustion and by the
change in direction  in $T_{eff}$ (see Figure~\ref{fig5} for the 2.22 $M_{\odot}$ case).
It corresponds to the sharp change in direction in the HR diagram in Figure~\ref{fig4}
(seen as a cusp for the 2.3 and 2.4 $M_{\odot}$ evolutionary tracks; for 2.2 and 2.5 $M_{\odot}$
cases the timesteps occured on either side of the time of central H exhaustion so the cusp was not resolved).
Additionally, this point is marked by a jump in H-burning CNO cycle luminosity (from roughly 70 to 90
$L_{\odot}$), while the surface luminosity does not change rapidly (nearly steady at 78 $L_{\odot}$)), 
rather the interior structure of the star is adjusting. 
He burning does not begin until 13 million years after this point in time. 
Between the time of central H exhaustion and time of He ignition the central density of the core increases from  $\sim100$ to $\sim 5\times 10^4$g/cm$^{-3}$.

We note that when HZ Her enters the stage of rapid radius expansion, the mass-loss rate onto 
Her X-1 will increase dramatically, by a factor of $\simeq 100$. 
This higher mass loss rate is super-Eddington, because the current mass transfer rate is at about 
0.1 of the Eddington rate. 
The likely outcome is unstable mass transfer, which happens for systems where mass transfer occurs 
from the more massive to less massive component, giving shrinkage of the orbit. 
This  instability occurs on a dynamical timescale and likely yields a common envelope system
(see Ivanova et al. 2013 
for a recent review of common envelope systems).
The common envelope system can leave behind a compact binary if the envelope is successfully ejected, 
consisting of the neutron star and the remaining white dwarf core of HZ Her. 
In this case the white dwarf would be a He white-dwarf, because He burning has not started 
in the core at the time of onset of the instability.
The neutron star can accrete from the white dwarf, giving an LMXB system, which could subsequently
evolve into a millisecond pulsar with a He white dwarf companion when the accretion ceases.

\subsubsection{New Limits on Mass of Her X-1, System Inclination and Distance}

New limits on $R$ and $T_{eff}$ for HZ Her have been determined here. 
By comparing with stellar evolution calculations, 
two allowed narrow mass ranges are found: 
2.15 to 2.20 $M_{\odot}$ if HZ Her has just past central hydrogen exhaustion,
and 2.35 to 2.45 $M_{\odot}$ if HZ Her has not yet reached central hydrogen exhaustion.
Since the duration of the evolutionary track in the allowed $R$ and $T_{eff}$ region 
is about 100 times less for the lower mass range (100kyr vs. 10Myr), 
the upper mass range is much more likely.

For either range, we can use the new limits on the mass of HZ Her,
together with the information summarized in  Table~\ref{tbl-massnew},
to narrow the original range of
mass of Her X-1 ($M_x$), system inclination, and $K_c$ (given in  Table~\ref{tbl-mass}).
We note that the mass ratio, $q=M_x/M_c$, can be computed using the standard formula
$q=K_c P_b/(2 \pi a_x sin(i))$,
valid for a circular orbit, which is a good approximation here given the very small orbital eccentricity. 
For the lower mass range, one finds $1.26<M_x/M_{\odot}<1.34$,  $82.1^{\circ}<i<90^{\circ}$ and
 99 km/s$<K_c<$102.8 km/s.
For the higher mass range, one obtains $1.47<M_x/M_{\odot}<1.71$,  $80^{\circ}<i<90^{\circ}$ and
 105.6 km/s$<K_c<$119 km/s.
The lower mass range for HZ Her yields a low-mass neutron star and the high mass range
yields a high-mass neutron star, but consistent with previously published estimates. 
However here, Table~\ref{tbl-massnew} specifies the dependence of neutron star mass on
inclination, $K_c$ and mass of HZ Her, so that an improvement in measurement of one of these quantities 
can yield a significant reduction in the allowed neutron star mass range.

Further, with the two new allowed mass ranges for HZ Her, the distance
limits (see Table~\ref{tbl-distance}) can be improved slightly.  
 For the lower mass range, which has $82.1^{\circ}<i<90^{\circ}$ and
  99 km/s$<K_c<$102.8 km/s, the upper and lower distance limits are 5.73 and 6.29 kpc. 
 For the higher mass range, which has  105.6 km/s$<K_c<$119 km/s, the distance
 limits are  5.87 and 6.78 kpc. 
 The uncertainties in inclination and $K_c$ both contribute to the distance uncertainty, 
 so that even with the constraint on the mass of HZ Her, the distance uncertainty is large. 

\section{Conclusions}

In this work, we analyzed X-ray eclipses of Her X-1 observed by the RXTE/PCA. 
A total of eight eclipse ingresses and egresses were found during Main High state 
which were free of absorption dips.
These eclipses were modeled to measure the radius of HZ Her and scale height of the atmosphere.
The scale height was found to be consistent with the previous determination by \citet{lea95}.
The radius measurements are the first accurate radius measurements of HZ Her, and have small
statistical uncertainty (1 part in 1000), 
with additional uncertainty caused by the unknown inclination and mass 
ratio of the HZ Her/Her X-1 binary system. 
The currently known orbital parameters, including an extreme range of inclinations (80 to 90$^\circ$) 
and range of $K_c$ (radial velocity amplitude of HZ Her, equivalent to $q$) 
are summarized in Table~\ref{tbl-orbit}. 
The implied semi-major axis and masses of HZ Her ($M_c$) and Her X-1($M_x$) are given 
in  Table~\ref{tbl-mass} .
We found that the radii derived from all eight eclipse observations agree with eachother, which
confirms that we obtained a reliable measurement of radius of HZ Her.
The resulting measured radius as a function of inclination and $K_c$ is given in Table~\ref{tbl-eclipseav}.

Next we considered the constraint on HZ Her from its optical spectrum observed during mid-eclipse,
when we see only the unheated face of HZ Her. 
We fitted the spectrum with Kurucz model atmospheres for a range of $T_{eff}$ and metallicity
with the surface gravity determined by the allowed masses (from orbital parameters) and allowed
radii (from X-ray eclipse).
We determined a 2$\sigma$ allowed range in $T_{eff}$ of 7720K to 7865K, and 
in metallicity of log($Z/Z_{\odot}$)=-0.27 to +0.03. 
With the observed continuum flux from HZ Her during mid-eclipse, we used the model surface flux and 
its 2$\sigma$ ranges to determine a best-fit distance to HZ Her and 2$\sigma$ upper and lower limits.
Because the radius inferred from eclipse depends on inclination and $K_c$, the results are presented
in  Table~\ref{tbl-distance} as a function of  inclination and $K_c$. 
The best-fit distance is 6.08 kpc, and the overall upper and lower limits are 5.73 kpc and 6.98 kpc
at the extremes of allowed  inclination and $K_c$.

Next we computed stellar evolutionary tracks for stars of various mass and solar metallicity
consistent with the narrow metallicity range from the spectral fits.
We required consistency of the model stars with both the temperature $T_{eff}$ and radius of HZ Her, which
has not been done before. 
The radius derived from X-ray eclipse depends on inclination and $K_c$, so each model mass has
a different constraint on radius. 
We calculated the constraints separately for each model mass. 
Only stars with mass in the one of two ranges of (2.15 to 2.20 $M_{\odot}$ 
and 2.35 to 2.45 $M_{\odot}$)  agree with the $T_{eff}$
and radius constraints at some point in their stellar evolution. 
Model stars in the low mass range are just past the time of central hydrogen exhaustion in
their evolution. 
Model stars in the higher mass range are not much younger than the time of central hydrogen exhaustion.
For the low mass range, the star is just entering the phase of rapid radius expansion, which should 
lead to rapid, unstable mass transfer. 
For the higher mass range, the star will enter this phase in about 40 Myr from now.
At the time of rapid radius expansion, HZ Her/Her X-1 will likely become a common envelope system, 
and then may emerge as an LMXB (neutron star with He white dwarf companion).

With the new mass limits on HZ Her from the stellar evolution calculations 
the range of allowed mass of the neutron star was determined.
For the lower mass range, the neutron star is low mass ($\simeq 1.34M_{\odot}$),
while for the higher mass range,  the neutron star is high mass ($\simeq 1.5-1.7M_{\odot}$).
Improvement in measurement of $K_c$ or system inclination can lead to significant improvement
in the limits on distance to Her X-1 and on neutron star mass.

\acknowledgments

D.A.L. thanks Prof. Didier Barret and Dr. Natalie Webb for hospitality at Institut de Recherche en Astrophysique et Planétologie (CNRS/UPS/OMP), Toulouse, France, where the final stages of this work were completed. 
We thank the referee, whose suggestions led to significant improvements in this paper.
The RXTE/PCA data were obtained from the RXTE Guest Observer Facility (GOF) at NASA's Goddard Space Flight Centre (GSFC).
This work was supported by the Natural Sciences and Engineering Research Council of Canada.

\clearpage


\clearpage

 \begin{figure}
 \epsscale{.80}
 \plotone{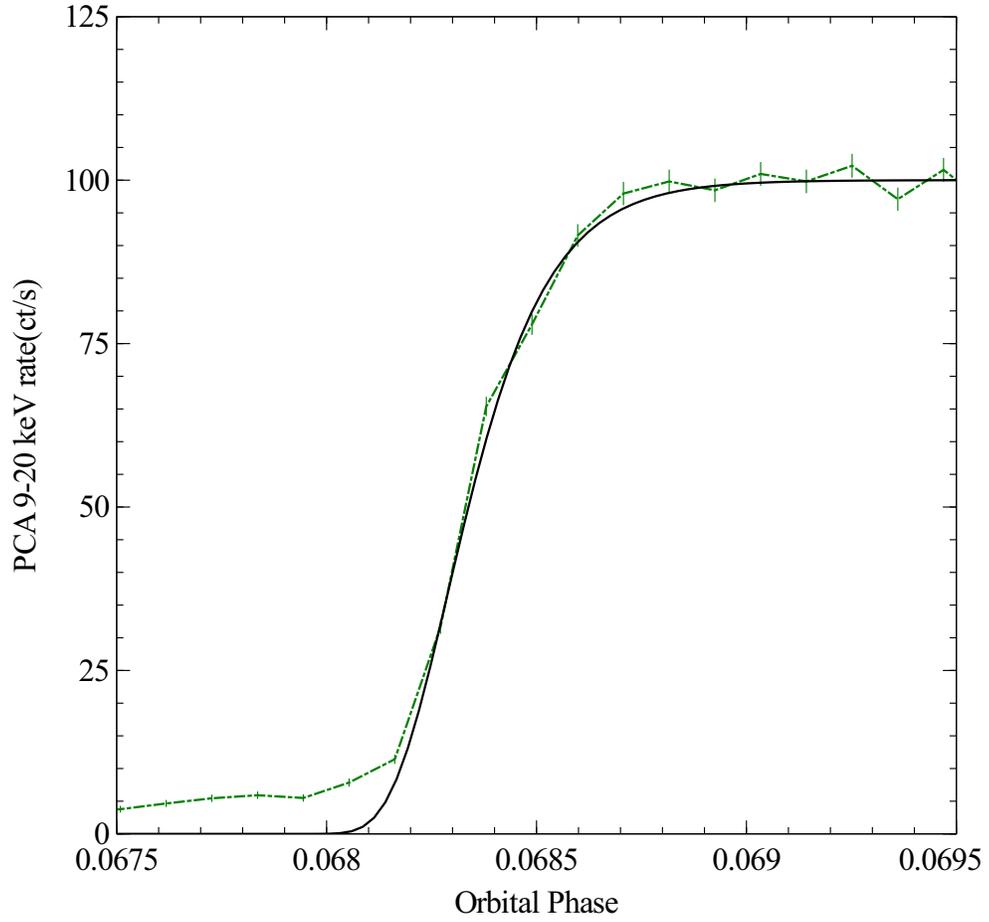}
 \caption{The model (black, solid line) and observed light curve for the eclipse egresse on MJD52597 (green, dash-dot line). See text for a description of the model.\label{fig2}}
 \end{figure}
 
 \clearpage

 \begin{figure}
 \epsscale{.80}
 \plotone{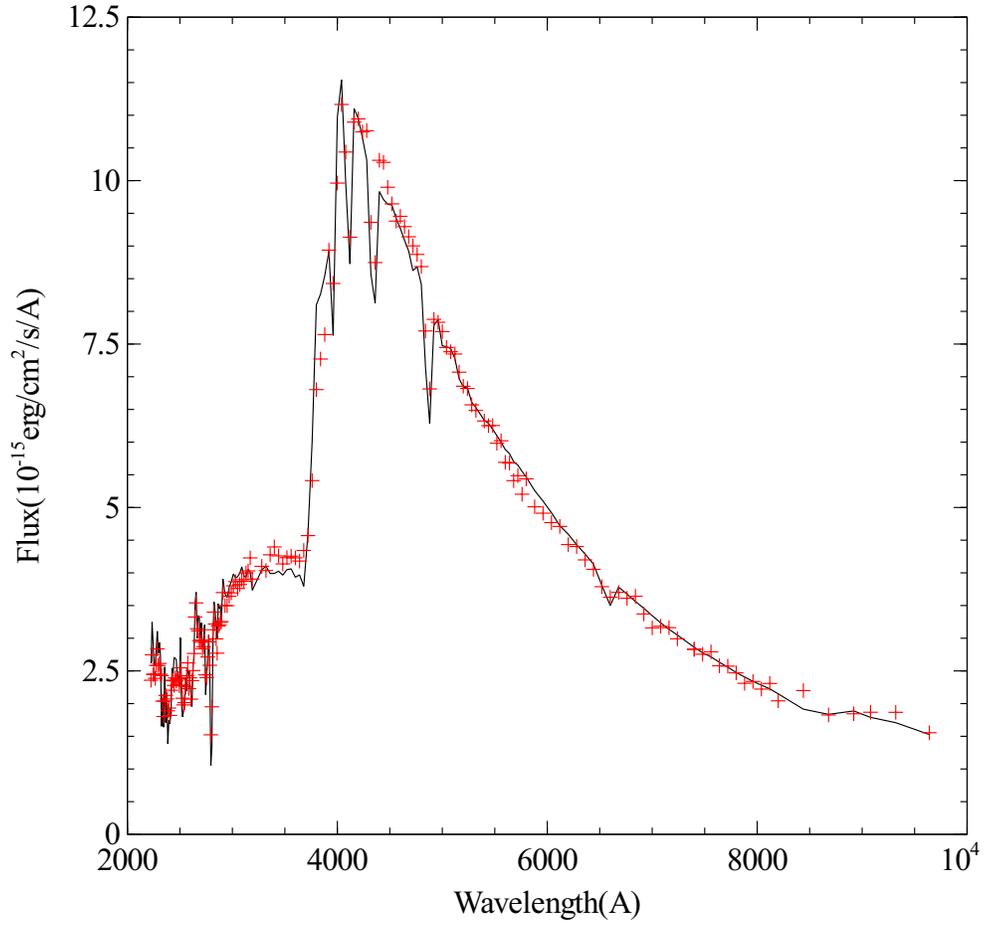}
 \caption{HZ Her observed spectrum during neutron star eclipse, at orbital phase 0,(red points) and best-fit Kurucz model atmosphere spectrum (black line) with $T_{eff}$=7800K, log($Z/Z_{\odot}$)=-0.1 and log(g)=3.5.\label{fig3}}
 \end{figure}
 
 \clearpage
 
  \begin{figure}
  \epsscale{.80}
  \plotone{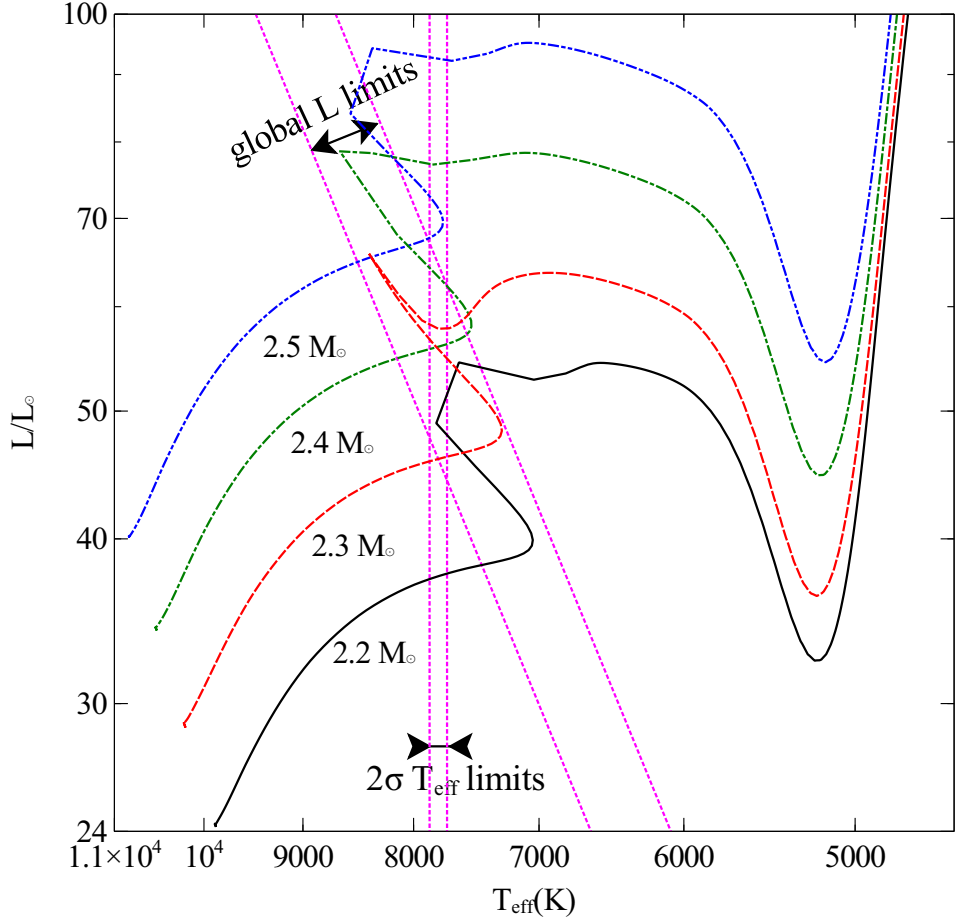}
  \caption{Stellar evolution tracks in the HR diagram for solar metallicity and initial masses (in
  units of $M_{\odot}$) of 2.2 (solid black line), 2.3(dash red line), 2.4(dash-dot green line) and 2.5(dash-dot-dot blue line). 
  This range in mass covers the allowed range by the measured orbital parameters.
  The upper and lower 2$\sigma$ limits to $T_{eff}$ are shown by the vertical dotted lines.
  Upper and lower limits to luminosity based on global upper and lower limits to the radius of HZ Her are shown by the sloped dotted lines.  \label{fig4}}
  \end{figure}
  
  \clearpage

    \begin{figure}
    \epsscale{.80}
    \plotone{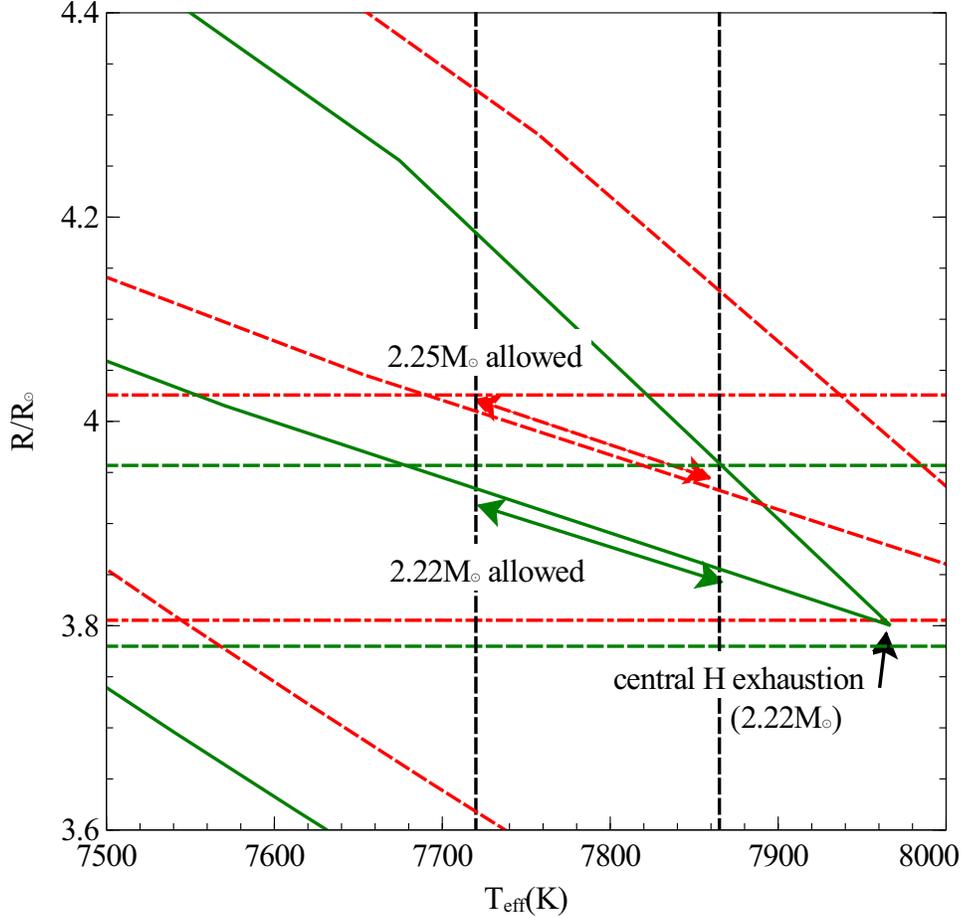}
    \caption{Evolution of 2.22 (green solid curve) and 2.25 (red dash curve)
    $M_{\odot}$ stars of solar metallicity in the $R$ vs. $T_{eff}$ plane. 
    The 2$\sigma$  limits on $T_{eff}$ from the fits to the optical spectrum of 
    HZ Her are plotted as the vertical black dash lines.
     The limits on $R$ from the fits to X-ray eclipse are dependent on mass of HZ Her: For 2.22 $M_{\odot}$, the upper and lower $R$ limits are the
     horizontal green dash lines, and for  2.25 $M_{\odot}$, the upper and lower $R$ limits are the horizontal red dash-dot lines.
     The 2.22 or 2.25 $M_{\odot}$ evolution tracks cross the 2.22 or 2.25 $M_{\odot}$ allowed 
     rectangles only for segments indicated by the arrows (green, solid arrow for 2.22 $M_{\odot}$,
     red, dash arrowfor 2.25 $M_{\odot}$).
     \label{fig5}}
    \end{figure}
    
    \clearpage






 \clearpage
 
 \begin{table}
 \begin{center}
 \caption{RXTE/PCA Main-High Eclipse Observations$^{a}$ \label{tbl-data}}
 \begin{tabular}{crrrr}
 \tableline\tableline
Type$^{b}$ & MJD & $\phi_{35}^{c}$ & $\phi_{orb,start}$ & $\phi_{orb,end}$ \\
 \tableline
 E & 50357 & 0.027 &.0617 & .0746 \\
 E & 50360 & 0.127 &.0617 & .0799 \\
 I & 50361 & 0.169 &.9200 & .9337 \\
 I & 50710 & 0.176 &.9201 & .9319 \\
 E & 52597 & 0.081 &.0645 & .0776 \\
 I & 52599 & 0.122 &.9200 & .9340 \\
 E & 52599 & 0.129 &.0670 & .0799 \\
 I & 52601 & 0.170 &.9200 & .9406 \\
 \tableline
 \end{tabular}
 \tablenotetext{a}{Start and end of continuous observations of each eclipse given.}
 \tablenotetext{b}{I for ingress, E for egress.}
 \tablenotetext{c}{$\phi_{35}$ is 35-day cycle phase.}
  \end{center}
  \end{table}
 
 \clearpage
 
 \begin{table}
  \begin{center}
  \caption{HZ Her/Her X-1 Orbital Parameters\label{tbl-orbit}}
  \begin{tabular}{crr}
      \tableline\tableline
      Parameter & Value(uncertainty) & Reference$^{a}$ \\
      \tableline
      $P_{orb}$ & 1.700167590(2)d &   (1) \\
      $\dot{P}_{orb}$ & $−4.85(13)\times 10^{-11}$ s/s   & (1)  \\
      $T_{\pi/2}$ &  MJD46359.871940(6)  & (1)  \\
      $a_x$sin(i) & 13.1831(4) lt-s &  (1)  \\ 
      eccentricity & $4.2(8)\times 10^{-4}$ & (1) \\
      $f_M$ & 0.85059(8) $M_{\odot}$ & (3)  \\ 
      $K_c$ & 109(10) km/s & (2) \\
      \tableline
      \end{tabular}
      \tablenotetext{a}{(1)= \citet{stau09}; (2)= \citet{rey97}; (3)= calculated here.}
  \end{center}
  \end{table}

    \clearpage
    
 \begin{table}
  \begin{center}
  \caption{Semi-major Axis$^{a}$ and Masses of HZ Her(M$_c$) and Her X-1 (M$_x$) \label{tbl-mass}}
  \begin{tabular}{crrrrrrrrr}
      \tableline\tableline
       $K_c$(km/s)  &  99  &   &   & 109   &   &   & 119 &   &   \\
       $q$ & 0.5856 &   &   & 0.6448 &   &   &  0.7039 &   &   \\
      Inclination & $a$ & $M_c$ & $M_{x}$ & $a$ & $M_c$ & $M_{x}$ & $a$ & $M_c$ & $M_{x}$ \\
      \tableline
      $80^{\circ}$ & 6.364  & 2.239 & 1.311 & 6.601   & 2.409  & 1.553 & 6.838 & 2.586 & 1.820   \\
      $82.5^{\circ}$ & 6.321  & 2.194 &  1.354 & 6.557 & 2.361  & 1.522 & 6.793 &  2.534 &  1.784 \\
      $85^{\circ}$ &  6.291  & 2.163 &  1.335  & 6.525 & 2.328 & 1.501 & 6.760 &  2.498 & 1.758 \\
      $87.5^{\circ}$ & 6.273  & 2.145 & 1.323 & 6.507 & 2.308 & 1.488 & 6.741 &  2.477 & 1.743 \\
      $90^{\circ}$ & 6.267 & 2.139 &  1.319 & 6.501 & 2.301 & 1.484 & 6.734 &  2.470&   1.738 \\
      \tableline
      \end{tabular}
      \tablenotetext{a}{Semi-major axis, a, is in units of $10^{11}$cm, masses are in $M_{\odot}$.}

  \end{center}
  \end{table}

    \clearpage
  
 \begin{table}
  \begin{center}
  \caption{Radii from X-ray Eclipse Models for 85$^\circ$ Inclination, $K_c$=109 km/s \label{tbl-eclipse85}}
  \begin{tabular}{crrr}
  \tableline\tableline
  Type$^{a}$ & MJD & $R_{tan}(\tau=1)$(cm)$^{b}$ & $H^{c}$(cm) \\
  \tableline
  E &50357 & $2.735\times10^{11}$ & $5\times10^{8}$ \\
  E &50360 & $2.734\times10^{11}$ & $5\times10^{8}$  \\
  I &50361 & $2.736\times10^{11}$ & $5\times10^{8}$ \\
  I &50710 & $2.735\times10^{11}$ & $5\times10^{8}$ \\
  E &52597 & $2.742\times10^{11}$ & $5\times10^{8}$ \\
  I &52599 & $2.736\times10^{11}$ & $5\times10^{8}$  \\
  E &52599 & $2.741\times10^{11}$ & $5\times10^{8}$  \\
  I &52601 & $2.734\times10^{11}$ & $5\times10^{8}$ \\
  \tableline
  \end{tabular}
  \tablenotetext{a}{I for ingress, E for egress.}
  \tablenotetext{b}{$R_{tan}(\tau=1)$ is the radius derived assuming a circular star, see text.}
  \tablenotetext{c}{$H$ is scaleheight of an exponential atmosphere.}
  \end{center}
  \end{table}
  
  \clearpage
  
 \begin{table}
  \begin{center}
  \caption{Radii, $R_{tan}(\tau=1)^{a}$, Averaged over Eclipses at Each Inclination \label{tbl-eclipseav}}
  \begin{tabular}{crrr}
  \tableline\tableline
     &  $K_c$=99 km/s &    $K_c$=109 km/s &    $K_c$=119 km/s \\
  Inclination &  & &  \\
  \tableline
  $80^{\circ}$ & 2.805   &  2.910 & 3.015 \\
  $82.5^{\circ}$ & 2.709  & 2.810 & 2.911 \\
  $85^{\circ}$ & 2.638 & 2.737 & 2.835 \\
  $87.5^{\circ}$ & 2.596 & 2.692 & 2.789 \\
  $90^{\circ}$ & 2.581 & 2.677 & 2.774 \\
  \tableline
  \end{tabular}
  \tablenotetext{a}{$R_{tan}(\tau=1)$ is given in units of $10^{11}$cm.  To obtain the equivalent
  radius $R_L$ of a Roche-lobe filling star with the same volume, multiply by a factor which depends on mass ratio and
  inclination (see text for details).}
  \end{center}
  \end{table}
  
  \clearpage

     \begin{table}
       \begin{center}
       \caption{$\chi^2$ for Fits$^{a}$ to HZ Her Near UV-Optical Spectrum
        \label{tbl-spectra}}
       \begin{tabular}{crrrrr}
       \tableline\tableline
          \tableline
         $T_{eff}$ & 7700K &  7750K & 7800K  & 7850K & 7900K \\
       \tableline
              log($Z/Z_{\odot}$) &     &   &  & &  \\
           -0.3 & 222.5 & 220.4 & 225.9 & 241.4 & 264.8 \\
           -0.2 & 221.0 &  213.9 & 214.2 & 224.7 & 244.1 \\
           -0.1 & 229.0  & 217.0 & 211.1 & 215.7 & 230.2 \\
            0.0 & 247.8 & 229.9 & 216.7 & 215.1 & 223.8 \\
           +0.1 & 278.1 & 254.0 & 233.0 & 224.7 & 226.9 \\
       \tableline
       \tableline
       \end{tabular}
       \tablenotetext{a}{Number of degrees of freedom is 182.}
       \end{center}
       \end{table}
       
       \clearpage

     \begin{table}
       \begin{center}
       \caption{2 $\sigma$ and 3 $\sigma$ limits to $T_{eff}$ and log($Z/Z_{\odot}$)
        \label{tbl-TZlimits}}
       \begin{tabular}{crrrrr}
       \tableline\tableline
         & lower limit &   lower limit    & best-fit & upper limit  & upper limit   \\
& 3$\sigma$ &  2$\sigma$  &  &  2$\sigma$  &  3$\sigma$   \\
     \tableline
       $T_{eff}$ & 7690K &  7720K & 7794K  & 7865K & 7895K \\
       \tableline
       log($Z/Z_{\odot}$) &  -0.33   & -0.27  & -0.12 & +0.03 & +0.09 \\
       \tableline
       \tableline
       \end{tabular}
       \end{center}
       \end{table}
       
       \clearpage    

  \begin{table}
   \begin{center}
   \caption{Distance$^{a}$ and 2$\sigma$ Upper and Lower Limits vs. Inclination
   and $K_c$ \label{tbl-distance}}
   \begin{tabular}{crrrrr}
   \tableline\tableline
   Inclination & $80^{\circ}$ &  $82.5^{\circ}$ & $85^{\circ}$  & $87.5^{\circ}$ & $90^{\circ}$ \\
   \tableline
   $K_c$=99 km/s &     &   &  & &  \\
   best-fit & 6.38    &  6.16 & 6.00 & 5.91 & 5.87 \\
   lower limit & 6.23  & 6.02 & 5.86 & 5.78 & 5.73 \\
   upper limit & 6.50 & 6.27 & 6.11 & 6.02 & 5.98 \\
      \tableline
 $K_c$=109 km/s &     &   &  & &  \\
   best-fit & 6.62    &  6.39 & 6.22 & 6.12 & 6.09 \\
   lower limit & 6.46  & 6.24 & 6.08 & 5.98 & 5.95 \\
   upper limit & 6.74 & 6.51 & 6.34 & 6.24 & 6.20 \\
      \tableline
   $K_c$=119 km/s &     &   &  & &  \\
      best-fit & 6.85   &  6.62 & 6.45 & 6.34 & 6.31 \\
      lower limit & 6.70  & 6.47 & 6.30 & 6.20 & 6.16 \\
      upper limit & 6.98 & 6.74 & 6.57 & 6.46 & 6.43 \\
   \tableline
   \end{tabular}
   \tablenotetext{a}{Distances are given in units of kpc.}
   \end{center}
   \end{table}
   
   \clearpage
   
 \begin{table}
  \begin{center}
  \caption{Inclination Range and Parameters as a Function of Mass of HZ Her \label{tbl-massnew}}
  \begin{tabular}{crrrrrrr}
      \tableline\tableline
        $M_c$  &  2.15 & 2.2  & 2.3 & 2.4  & 2.5 & 2.55      \\
      \tableline
      Minimum & & & & & & \\
      $i_{min}$ &  $86.74^{\circ}$  & $82.11^{\circ}$  & $80^{\circ}$ & $80^{\circ}$ & $80^{\circ}$ & $80^{\circ}$    \\
      $K_{c,min}$ & 99.0 & 99.0 & 102.6 & 108.5 & 114.2 & 117.0  \\
      $a_{min}$ & 6.279	 & 6.327 &	6.450 &	6.588 &	6.724 &	6.791 \\
      $M_{x,min}$ & 1.260 &	1.289 &	1.396 &	1.540 &	1.689 &	1.765  \\
      $R(i_{min})$ & 2.605	& 2.718 &	2.843 &	2.904 &	2.964 &	2.971   \\
        & & & & & & & \\
      Maximum & & & & & & \\
      $i_{max}$ &  $90^{\circ}$  & $90^{\circ}$  & $90^{\circ}$ & $90^{\circ}$ & $84.82^{\circ}$ & $81.63^{\circ}$    \\
      $K_{c,max}$ & 99.7 &	102.8 &	108.9 &	114.9 &	119.0 &	119.0 \\
      $a_{max}$  & 6.284 &	6.356 &	6.499 &	6.639 &	6.776 &	6.843  \\
      $M_{x,max}$ & 1.268 &	1.338 &	1.482 &	1.631 &	1.786 &	1.865  \\
      $R(i_{max})$ & 2.582 &	2.618 &	2.677 &	2.734 &	2.839 &	2.943  \\
      \tableline
      \end{tabular}
      \tablenotetext{a}{Semi-major axis, a, and radius $R$ are in units of $10^{11}$cm, masses are in $M_{\odot}$,
       $K_{c}$ in units of km/s.}

  \end{center}
  \end{table}

    \clearpage


\begin{thebibliography}{}
\bibitem[Anderson et al.(1994)]{and94} Anderson, S.~F., 
    Wachter, S., Margon, B., et al.\ 1994, \apj, 436, 319 
\bibitem[Boroson et al. (1997)]{bor97}
    Boroson, B., Blair, W. P., Davidsen, A. F., et al.  1997, \apj, 491, 903
\bibitem[Cheng et al. (1995)]{cheng95} 
    Cheng, F. H., Vrtilek, S. D. \& Raymond, J. C. 1995, \apj, 452, 825
\bibitem[Choi et al. (1994)]{cho94}
    Choi, C., Dotani, T., Nagase, F., Makino, F.,  Deeter, J. \& 
    Min, K. 1994, ApJ, 427, 400
\bibitem[Choi et al. (1997)]{cho97}  Choi, C., Seon, K., Dotani, T.,
    \& Nagase, F. 1997, \apj, 476, L81 
\bibitem[Crosa \& Boynton (1980)]{cro80} Crosa, L., \& Boynton, P. E. 1980,  \apj, 235, 999
\bibitem[Dal Fiume et al. (1998)]{dal98}
    Dal Fiume, D., Orlandini, M., Cusumano, G., et al. 1998, \aap, 329, L41
\bibitem[Deeter et al. (1976)]{dee76}
    Deeter, J., Crosa, L., Gerend, D., \& Boynton, P. 1976, \apj, 206, 861
\bibitem[Eggleton(1983)]{egg83} Eggleton, P.~P.\ 1983, \apj, 268, 368 
\bibitem[Eggleton(1971)]{egg71} Eggleton, P.~P.\ 1971, 
\mnras, 151, 351 
\bibitem[Gerend and Boynton (1976)]{ger76}
    Gerend, D., \& Boynton, P. 1976, \apj, 209, 652
\bibitem[Igna (2010)]{inga10}  Igna, C.D, 2010, PhD Thesis, University of Calgary
\bibitem[Igna \& Leahy(2011)]{ign11} Igna, C.~D., \& Leahy, D.~A.\ 2011, \mnras, 418, 2283
\bibitem[Igna \& Leahy(2012)]{ign12} Igna, C.~D., \& Leahy, D.~A.\ 2012, \mnras, 425, 8 
\bibitem[Ivanova et al.(2013)]{iva2013} Ivanova, N., Justham, S., Chen, X., et al.\ 2013, \aapr, 21, 59 
\bibitem[Ji et al.(2009)]{ji09} Ji, L.
    Schulz, N., Nowak, M., Marshall, H.~L., \& Kallman, T. 2009, ApJ, 700, 977
\bibitem[Jimenez-Garate et al.(2005)]{jim005} Jimenez-Garate, 
    M.~A., Raymond, J.~C., Liedahl, D.~A., 
    \& Hailey, C.~J.\ 2005, \apj, 625, 931 
\bibitem[Klochkov et al.(2009)]{klo09} Klochkov, D., 
    Staubert, R., Postnov, K., Shakura, N., \& Santangelo, A. 2009, A\& A, 506, 1261 
\bibitem[Kurucz(1993)]{Kur93} Kurucz, R.~L.\ 1993, IAU 
    Colloq.~138: Peculiar versus Normal Phenomena in A-type and Related Stars, 
    44, 87 
\bibitem[Leahy \& Yoshida(1995)]{lea95} Leahy, D.~A., \& Yoshida, A.\ 1995, \mnras, 276, 607
\bibitem[Leahy (1995a)]{lea95a}
    Leahy, D.A. 1995a, \apj, 450, 339
\bibitem[Leahy (1995b)]{lea95b}
    Leahy, D.A. 1995b, A{\&}ASS, 113, 21
\bibitem[Leahy (1997)]{lea97}
    Leahy, D.A. 1997, MNRAS, 287, 622
\bibitem[Leahy \& Scott (1998)]{lea98}  
    Leahy, D., \& Scott, D. M. 1998,  \apj, 503. L63 
\bibitem[Leahy (2002)]{lea02}
    Leahy, D.A. 2002, MNRAS, 334, 847
\bibitem[Leahy (2003)]{lea03}
    Leahy, D.A. 2003, MNRAS, 342, 446
\bibitem[Leahy (2004)]{lea04}
    Leahy, D.A. 2004, MNRAS, 348, 932
\bibitem[Leahy(2004b)]{lea04b} Leahy, D.~A.\ 2004b, \apj, 613, 517     
\bibitem[Leahy \& Igna (2010)]{lea10} Leahy, D.~A., \& Igna, C. 2010, \apj, 713, 318 
\bibitem[Leahy \& Igna(2011)]{lea11} Leahy, D.~A., \& Igna, C.\ 2011, \apj, 736, 74 
    
\bibitem[McCray et al. (1982)]{mcc82}
    McCray, R., Shull, M., Boynton, P., Deeter, J., Holt, S., White, N. 
    1982 \apj 262, 301 
\bibitem[Oke(1976)]{oke76} Oke, J.~B.\ 1976, \apj, 209, 547 
\bibitem[Oosterbroek et al. (1998)]{oos98}
    Oosterbroek, T., Parmar, A., Martin, D., \& Lammers, U. 1998 \aap 329, L41 
\bibitem[Oosterbroek et al. (2000)]{oos00}
    Oosterbroek, T., et al. 2000, A\&A, 353, 575 
\bibitem[Reynolds et al. (1997)]{rey97}
    Reynolds, A., Quaintrell, H., Still, M., Roche, P., Chakrabarty, D. \&
    Levine, S. 1997, \mnras, 288, 43 
\bibitem[Scott \& Leahy (1999)]{sco99} Scott, D. M., \& Leahy, D. 1999, 
    \apj, 510, 974
\bibitem[Scott et al. (2000)]{sco00} Scott, D.~M., Leahy, 
    D.~A., \& Wilson, R.~B. 2000, \apj, 539, 392 
\bibitem[Shakura et al. (1998)]{sha98} Shakura, N., Postnov, K., \&
    Prokhorov, M. 1998, \aap, 331 , L37
\bibitem[Staubert et 
    al.(2009)]{stau09} Staubert, R., Klochkov, D., \& Wilms, J.\ 2009, \aap, 500, 883 
\bibitem[Shulman et al. (1975)]{shul75} Shulman, S., Friedman, H., Fritz, G.,
    Henry, R. C., \& Yentis, D. J. 1975, \apjl, 199, L101
\bibitem[Still et al. (1997)]{sti97}
    Still, M., Quaintrell, H., Roche, P., Reynolds, A. 1997, \mnras, 292, 52 
\bibitem[Voloshina et al. (1990)]{vol90}
    Voloshina, I., Lyutyi, V., \& Sheffer, K. 1990,
    Sov. Astron. Lett. 16, 257
\bibitem[Vrtilek \& Cheng (1996)]{vrt96}
    Vrtilek, S. D., \& Cheng, F. H.  1996, \apj, 465, 915
    
\end{thebibliography}
\end{document}